\documentstyle[amsfonts,epsfig,12pt]{article}

\def\mypagenumber{1}

\def\myend{\end{document}}

\normalsize

\newcounter{sxn}

\newcounter{axn}

\date{}

\newdimen\mybaselineskip
\newcommand{\beeq}{\begin{equation}}
\newcommand{\eneq}{\end{equation}}
\newcommand{\be}{\begin{eqnarray}}
\newcommand{\ee}{\end{eqnarray}}
\newcommand{\bpic}{\begin{picture}}
\newcommand{\epic}{\end{picture}}

\def\dd{\partial}
\def\la{\raise.16ex\hbox{$\langle$} \, }
\def\ra{\, \raise.16ex\hbox{$\rangle$} }

\def\psibar{ \psi \kern-.65em\raise.6em\hbox{$-$} }
\def\mbar{ m \kern-.78em\raise.4em\hbox{$-$}\lower.4em\hbox{} }

\def\ep{\epsilon}

\def\n@space{\nulldelimiterspace=0pt \mathsurround=0pt }
\def\huge#1{{\hbox{$\left#1\vbox to 20.5pt{}\right.\n@space$}}}

\def\myskip{\noalign{\kern 8pt}}
\def\myeqspace{\noalign{\kern 10pt}}

\def\boxit#1{$\vcenter{\hrule\hbox{\vrule\kern3pt
    \vbox{\kern3pt\hbox{#1}\kern3pt}\kern3pt\vrule}\hrule}$}
\def\bigbox#1{$\vcenter{\hrule\hbox{\vrule\kern5pt
     \vbox{\kern5pt\hbox{#1}\kern5pt}\kern5pt\vrule}\hrule}$}

\def\ignore#1{{}}

\begin{document}

\bibliographystyle{unsrt}
\footskip 1.0cm

\thispagestyle{empty}
\setcounter{page}{\mypagenumber}

             
\begin{flushright}{
BRX-TH-496\\}

\end{flushright}

\vspace{2.5cm}
\begin{center}
{\LARGE \bf {Yang-Mills Solutions on Euclidean Schwarzschild  
Space }}\\ 
\vskip 1 cm
{\large{Bayram  Tekin  }}\footnote{e-mail:~
 tekin@brandeis.edu}\\
\vspace{.5cm}
{\it Department of  Physics, Brandeis University, Waltham,02454  MA,
USA}\\

\end{center}

\vspace*{2.5cm}


\begin{abstract}
\baselineskip=18pt

\end{abstract}

We show that the apparently 
periodic Charap-Duff Yang-Mills `instantons'  
in time-compactified Euclidean Schwarzschild space are actually time 
independent. For these solutions, the 
Yang-Mills potential is constant along the 
time direction (no barrier) and therefore, there is no 
tunneling. We also demonstrate that the solutions found to date are
three dimensional monopoles and dyons. We conjecture that there are no
time-dependent solutions in the Euclidean Schwarzschild background.

\vfill
 
\newpage



\normalsize
\baselineskip=22pt plus 1pt minus 1pt
\parindent=25pt
\vskip 2 cm

Studies \cite{cd1,cd2} of Yang-Mills (YM) instantons coupled to 
Einstein's gravity started not long after their discovery in flat space 
\cite{polyakov}, but this subject has not received proper interest since. 
Lorentzian Einstein-YM system  
holds quite a number of surprises-one 
of which is a soliton solution \cite{bk} which is absent both  
in pure gravity and in pure YM theory \cite{deser1}. 
In Euclidean space, the 
energy momentum tensor of self-dual YM solutions vanishes and therefore
gravity is not disturbed by the presence of instantons but 
does affect them,~\footnote{A similar situation arises in
three dimensions for a different reason,  
the imaginary nature of the Chern-Simons action in Euclidean space :
there are 
Chern-Simons gravity-YM-Higgs 
`gravitating monopole-instanton' 
solutions which also do not curve space \cite{ferstl}.}
in a number of different ways : in particular 
it can change the space-time topology and it can bring a scale 
(or multiple scales). 
If space is conformally flat, the
self-duality equations of flat-space YM theory are intact and the flat space 
BPST instantons are formally solutions. 
But this is too naive: one has to take into 
account the existence of the horizons and the global topology of the 
space-time. Euclidean de Sitter is 
a nice example in this context: there is a horizon and the time is 
compactified. In the 
literature \cite{chak} one can find {\it{static}} three dimensional 
solutions but apparently no fully four-dimensional instanton solutions. 

Here we shall take the background to be Euclidean Schwarzschild space. 
YM theory in this background was studied long ago  by Charap and 
Duff (CD) \cite{cd1,cd2}, who found self-dual solutions. However the 
physical meaning of the solutions of CD  
has not hitherto been resolved. CD acknowledge that their `instanton' 
solutions, being periodic in Euclidean time, do not allow a tunneling 
interpretation between vacua.  
This work started with the intention to show  
that periodic instanton solutions in a curved background can be given 
a normal tunneling interpretation in the same spirit as the caloron 
solutions \cite{hk} of YM theory on $S^1\times R^3$. 
As demonstrated in  \cite{dunne}, the caloron gauge field, initially
constructed to be explicitly periodic, is no longer periodic in the Weyl 
gauge ( $ A_0^a = 0$ ), which is more suitable for the Hamiltonian
processes, such as  tunneling. But, unlike the calorons, 
the apparently time-periodic YM solutions 
of CD in Euclidean Schwarzschild turned out to be actually 
{\it{time-independent}}, when 
looked at the proper gauge. As we shall show, these solutions are more 
like BPS monopoles than 
instantons. In fact, the YM potential is constant along the time 
direction for these solutions and there is no barrier to tunnel. 

The Euclidean Schwarzschild space,
in Schwarzschild and Kruskal coordinates respectively, is  
\be
ds^2 &=& H(r) \, dt^2 + {1\over H(r)} dr^2   
 + r^2 (d\theta^2 + \sin^2 \theta \, d\phi^2), \hskip 0.3 cm  
H(r) =  1 - {2M \over r}, 
\label{metric1} \\
ds^2 &=& {32 M^3\over r}\mbox{exp}(- {r\over 2M})( dz^2 + dy^2) + 
r^2 (d\theta^2 + \sin^2 \theta \, d\phi^2), 
\label{kruskal}
\ee
where  
\be
z^2 + y^2 = ( {r\over 2M} - 1)\mbox{exp}({r\over 2M}), \hskip 2 cm  
{y - iz\over y+iz} = 
\mbox{exp} {-it\over 2M}. 
\label{kru}
\ee
The Euclidean time can be compactified to remove the singularity at the origin.
[We set the gravitational 
constant $G$ and the Yang-Mills coupling constant $g_{YM}$ to unity: 
$G = g_{YM} = 1$.]
\be
t \rightarrow t + \beta  ,\hskip 1 cm \beta = 8\pi M, \hskip  1 cm  r \ge 2 M.
\ee
However, once compactified, the  $M \rightarrow 0$ 
limit is tricky, and formally  
it yields a three instead of four dimensional 
Euclidean space. (This suggests 
that YM instantons in this background, if they existed at all, would not limit 
to the flat space $4D$ BPST instanton for $M \rightarrow 0$. )

We take $SU(2)$ YM theory and  adopt the spherically symmetric instanton 
ansatz for the YM gauge connection,
\be
A = {1\over 2 }\tau^a A_{\mu}^a dx^\mu =  {\tau^a\over 2} \Bigg\{ 
{A_0}  {x_a\over r}  dt + {A_1}  {x_a  x_k\over r^2} dx_k +{\phi_1\over r} 
 \bigg(\delta_{a k} - {x_a x_k\over r^2} \bigg)  dx_k  
+ \ep_{a kl} {\phi_2 -1 \over r^2} { x_k} dx_l \Bigg\}, 
\label{ym}
\ee
where $\tau^a$ are Pauli matrices. We should mention that Charap-Duff 
\cite{cd2}  have only three unknown functions in their ansatz,as  
they work with an already gauge-fixed instanton 
ansatz ( $x^j\, A^a_j = A_1 = 0$ ) from the start. On the  
other hand we have not fixed the gauge yet. This will be a 
crucial point for the physical interpretation of the solutions.

Our task is to solve the self-duality equations, $F = \pm *F$, but there are 
at least one apparent and one serious obstruction 
to the existence of non-trivial instanton solutions. The apparent one is that
naive topological considerations seem to imply  
that there will be no  non-trivial solutions of equations of motion.
The topology of the Euclidean Schwarzschild space is $R^2\times S^2$.  
At spatial infinity ($r \rightarrow \infty $), the space has the topology
$S_{\beta}^1\times S_\infty^2 $ ( we add the point $t =0$ to the time direction).
The relevant homotopy groups vanish ( $\Pi_2(SU(2)) = 0$ 
and  $\Pi_1(SU(2)) = 0$); considered as a map, 
$ A: S^1\times S^2 \rightarrow SU(2)$ cannot have 
a winding number. 
But this  argument is not sufficient to  prove that there are  
no non-trivial solutions, 
the existence of calorons in the same (asymptotic) topology being 
 a counterexample. There {\it{can}} indeed be 
non-trivial maps between $S^1\times S^2$ and $SU(2)$. For explicit construction of these maps in the finite
temperature gauge theory context,  
we refer the reader to  \cite{pisarski} and \cite{gerald}. 

Another way to avoid this topological 
`simplicity'  is to consider solutions which have non-trivial holonomy for the Polyakov loop at spatial infinity: 
${\cal{P}}\mbox{exp}\int_0^\beta A_0 dt \ne \pm 1$. This will effectively break the gauge group to 
$U(1)$ and non-trivial homotopy will guarantee the solutions. But more generally we can rely 
on the cohomology arguments for the existence of non-trivial solutions. 
Namely we need the
topological 
charge to be non-vanishing, expressed as an integral over the boundary : 
$Q \sim \int_M tr{F\wedge F} = \int_M d K = \int_{S^2\times S^1} K $. 
A non-exact but closed 3-from $K$ 
can be  constructed since the relevant cohomology groups are non-trivial: 
$H^1(S^1,R) = R$ and $H^2(S^2,R) = R$. This 
argument alone does not imply the quantization of the topological 
charge. This is obtained by the requirement of the 
single valuedness of the path-integral $< \mbox{exp} (i Q) >$.

The more serious obstruction for the existence of arbitrary size 
instanton solutions is
the existence of a  scale,  $2 M$. It is clear that YM theory in the Schwarzschild background is 
not scale invariant and the best one can do is to look 
for `constrained instantons' as in \cite{affleck}. 
These are approximate solutions with scales restricted to a domain, 
usually given by the scale in the theory. At the end of 
this paper, we will argue that 
Euclidean Schwarzschild space do not allow such solutions. 

We next show that the solutions 
reported in the literature are 3D monopoles rather than 4D instantons. 
  
The gauge field strength , 
$F = dA - iA\wedge A$,  
is computed to be \cite{witten,hosotani}

\be
F&=&
{1\over 2}\bigg\{
(\dot{A_1}-A_0^{\prime})\tau_3 dt\wedge dr
+\Big[ (\dot{\phi_2}-A_0\phi_1)\tau_1 + (\dot{\phi_1}+A_0 \phi_2)\tau_2\Big]
dt\wedge d\theta \cr
\noalign{\kern 8pt}
&&\hskip 1cm
-\Big[(\dot{\phi_1}+A_0\phi_2 )\tau_1+(-\dot{\phi_2} + A_0\phi_1 )\tau_2 \Big]
dt\wedge\sin\theta d\phi\cr
\noalign{\kern 8pt}
&&\hskip 1cm
+\Big[ (\phi_2^{\prime}-A_1\phi_1)\tau_1
   +(\phi_1^{\prime}+ A_1\phi_2 )\tau_2\Big] dr\wedge d\theta \cr
\noalign{\kern 8pt}
&&\hskip 1cm
+ \Big[(-\phi_1^{\prime}-A_1 \phi_2)\tau_1 + 
(\phi_2^{\prime}-A_1\phi_1)\tau_2\Big]   dr\wedge\sin\theta d\phi \cr
\noalign{\kern 8pt}
&&\hskip 1cm
-(1-\phi_1^2 -\phi_2^2)\tau_3 d\theta\wedge\sin\theta d\phi \bigg\} ~~.
\label{field1}
\ee
Here $\phi^\prime$ and  $\dot{\phi}$ denote derivative with respect to $r$
and $t$ respectively.
The dual field strength, in the Euclidean Schwarzschild background, is
\be
*F&=&
{1\over 2}\bigg\{-{1\over r^2}(1-\phi_1^2 -\phi_2^2)\tau_3 dt\wedge dr
+H(r)\Big[ (\phi_1^{\prime}+A_1\phi_2)\tau_1 +
   (-\phi_2^{\prime}+ A_1\phi_1 )\tau_2\Big]
dt\wedge d\theta \cr
\noalign{\kern 8pt}
&&\hskip 1cm
+ H(r) \Big[(\phi_2^{\prime}-A_1 \phi_1)\tau_1 + 
(\phi_1^{\prime}+A_1\phi_2)\tau_2\Big]
dt\wedge\sin\theta d\phi\cr
\noalign{\kern 8pt}
&&\hskip 1cm
-{1\over H(r)}\Big[ (\dot{\phi_1}+A_0\phi_2 )\tau_1+(-\dot{\phi_2}+ A_0\phi_1 )\tau_2 \Big]
 dr\wedge d\theta \cr
\noalign{\kern 8pt}
&&\hskip 1cm
+ -{1\over H(r)}\Big[ (\dot{\phi_2} -A_0\phi_1)\tau_1
   +(\dot{\phi_1} + A_0\phi_2 )\tau_2\Big]    dr\wedge\sin\theta d\phi \cr
\noalign{\kern 8pt}
&&\hskip 1cm
+r^2(\dot{A_1}-A_0^{\prime})\tau_3 d\theta\wedge\sin\theta d\phi \bigg\} ~~.
\label{field2}
\ee

Before we look at the equations of motion, let us show that there is a nice
dimensional reduction of the  
four dimensional YM action to a 2D  abelian Higgs model in a curved background. 
\be
S_{YM}&=& \int_{M} \mbox{tr} F\wedge *F
\nonumber \\ 
&=& 8\pi\int_{\Sigma} d^2 x \, \sqrt{h} \Bigg\{ 
{1\over 2}h^{\mu\nu}D_\mu \varphi_i\,D_\nu \varphi_i 
+{1\over 8}h^{\mu\alpha}\,h^{\nu\beta}  F_{\mu\nu} F_{\alpha \beta} +
{1\over {4}}(1 -\varphi_i^2)^2  \ \Bigg \}
\label{abelian}
\ee
Space-time indices refer to $(t, r)$ only and $i, j = (1,2)$. 
Where $F_{\mu \nu}= \dd_\mu A_\nu - \dd_\nu A_\mu$ and 
$D_\mu \varphi_i= \dd_\mu \varphi_i +\epsilon_{ij} A_\mu \varphi_j$ are the 
2 dimensional abelian field strength and covariant derivative 
respectively. $\Sigma$ 
is a semi-infinite strip in the upper-half plane with the following metric
\be
ds^2 = h_{\mu\nu} dx^\mu\, dx^\nu =  {H(r)\over r^2} dt^2 + {1\over r^2 H(r)} dr^2.
\label{metric2}
\ee
This reduction of course follows from the conformal invariance of 
the Yang-Mills 
action. One can simply pull out a factor of $r^2$ from the metric 
(\ref{metric1}). The result (\ref{abelian}) merely generalizes 
that of the flat space case \cite{witten}: four dimensional Yang-Mills theory on $R^4$, for 
spherically symmetric solutions, reduces to the two dimensional abelian-Higgs model
on the upper half-plane with the Poincare metric $ds^2 = r^{-2}( dt^2 + dr^2)$.

Now let us study the self-duality equations $F= \pm *F $, expressed as 

\be
&& \dot{A_1}-  A_0^{\prime} =  \mp {1\over r^2} (1-\phi_1^2 -\phi_2^2), \cr
&&  \dot{\phi_2}- A_0\phi_1 = \pm H(r) (\phi_1^{\prime}+A_1\phi_2 ), \cr
&&  \dot{\phi_1}+ A_0 \phi_2 = \mp H(r)(\phi_2^{\prime}-A_1 \phi_1) 
\label{selfdual1}
\ee
Both the equations of motion and the action are invariant under the 
non-abelian gauge transformations 
( $U(\vec{x},t) = \mbox{exp}-i{f(r,t)\over 2}\hat{x}\cdot \vec{\tau}$)which transform 
the ansatz functions in the following way,
\be
\tilde{A_0} &=& A_0 + \dot f \cr
\tilde{A_1} &=& A_1 + f' \cr
\tilde{\phi_2} &=& +\phi_2 \cos f + \phi_1 \sin f \cr
\tilde{\phi_1} &=&  -\phi_2 \sin f+ \phi_1 \cos f .
\label{gauge2}
\ee
In flat space, $M =0$ and $H(r) = 1$, the equations (\ref{selfdual1}), 
augmented with a gauge condition, are integrable and were solved
in \cite{witten} to obtain n-instanton 
solutions located on the t-axis at arbitrary 
locations and with arbitrary sizes. 
[Another interesting fact about these equations was noted   
in \cite{comtet} and \cite{tekin}: (\ref{selfdual1})
corresponds to extremal  surfaces in the  $R^{(2,1)}$, for $H(r) =1$. For 
generic $H(r)$, the equations (\ref{selfdual1}) can be shown to reduce to a 
problem of finding a (non-extremal) surface whose Gaussian curvature is a 
function of H(r) \cite{barti}. ] 

Now let us look at non-zero $M$. The charge 1 solution given by \cite{cd2} is
\be
&& A_1 = 0, \hskip 4 cm  A_0 = \lambda - {M\over r^2} , \cr
&&\phi_1 = (1- {2M\over r})^{1\over 2}\cos ( {\lambda t} + \omega_0), 
\hskip 1 cm  \phi_2 = (1- {2M\over r})^{1\over 2}\sin ( {\lambda t} + \omega_0)
\label{duff}
\ee    
This solution will be smooth  $ r\in [ 2M, \infty )$ and $t \in [0, 8\pi M]$,   
if one sets $\lambda = 1/4M$. The non-abelian gauge field is periodic  
( $A_{\mu}^a(r,t=0)= A_{u}^a(r, t= 8\pi M)$ ). $\omega_0$ is an  
integration constant which denotes 
the angular location of the instanton in the 
time interval and without loss of generality we choose $\omega_0 = \pi/2$.  
Observe that there is no (arbitrary) size for the solution. 
In this periodic gauge, the flat space limit ( $M \rightarrow 0$) does not seem to be 
well-defined. But it is clear that we can gauge transform this solution 
to a time independent solution using (\ref{gauge2}), with a gauge transformation 
function  $f(r,t) = - \lambda t$. Then the solution will be
 
\be
&& A_1 = 0, \hskip 1.5 cm  A_0 = - {M\over r^2} , \cr
&&\phi_1 = 0, 
\hskip 1 cm  \phi_2 = (1- {2M\over r})^{1\over 2}
\label{du2}
\ee    
This solution has an artificial (gauge) singularity at $r = 2M$ and this  is the same solution 
as the one CD obtained from the spin connection in their first paper on the subject \cite{cd1}.
We can show that it is a  charge 1 monopole, by associating  an abelian field strength 
through  't Hooft's \cite{thooft} definition  [and also see Abbott-Deser 
\cite{ad} ] 
\be
f_{\mu \nu} = \partial_\mu ( \hat{A}_0^a A_\nu^a )- 
\partial_\nu ( \hat{A}_0^a A_\mu^a) + \epsilon^{abc}\hat{A}^a_0 
\partial_{\mu}  \hat{A}^b_0 \partial_\nu \hat{A}^c_0 \
\label{ab}
\ee
where  $\hat{A}^a_0 = A^a_0/ \sqrt{(A_0^a A_0^a)}$. 
\be
f_{0 i} = { x_i A_0^\prime\over r}, \hskip 1 cm  f_{ij} = {\epsilon_{ija} x^a \over r^2}
\label{electmag}
\ee
The electric field $E_i = f_{0 i}$ decays rapidly and so the electric charge of the solution is 
zero but the magnetic charge is 1. Thus, as promised, we have demonstrated that the solution of
\cite{cd2} is time independent and corresponds to a monopole. Another way to see this is to 
look at the gauge invariant  Yang-Mills potential. It reads
\begin{eqnarray}
V(t)=2\pi \int_{2M}^\infty dr\left\{2H(r)( \phi_1^{\prime}+A_1
\phi_2)^2+2H(r) (\phi_2^\prime - A_1 \phi_1)^2+\frac{1}{r^2} (1-\phi_1^2
-\phi_2^2)^2\right\}
\label{pot}
\end{eqnarray}
For (\ref{duff}), we have
\be
V =  {\pi\over 2 M}.
\label{pot2}
\ee
It is time independent and so there is no barrier to tunnel unlike the flat-space BPST
instanton or the time-periodic caloron \cite{dunne}. 
Its kinetic energy is
\be
E_{kinetic}(t) = 2\pi \int_{2M}^\infty dr\left\{ {2\over H(r)} ( \dot{\phi_1}+ A_0
\phi_2)^2+{2\over H(r)} (\dot{\phi_2} - A_0 \phi_1)^2+ r^2 ( \dot{A_1} - A_0^\prime)^2 \right\},
\label{kin}
\ee
which is computed to be
\be
E_{kinetic}(t) =  {\pi\over 2 M}. 
\label{kin2}
\ee
We have seen that, even though the monopole has an associated 
electric field, its flux at infinity is zero. 
The `mass' of the monopole can be defined  as the sum of the kinetic and the potential energies: namely 
$\pi/ M$. In flat space the monopole action ($\int \mbox{ mass}\, dt $) is divergent, but here, 
because of the compactness of the time dimension, its action is  $ 8 \pi^2 $. This is numerically  
same as the topological charge 1 instanton action, but as argued above, the solution at hand is 
a BPS monopole.

We can also give a suggestive argument of how to interpret the mass of the monopole.  
Needless to say, $A^a_0$ plays the role of an adjoint Higgs field, as is already clear from the definition
of the abelian field strength (\ref{ab}). The flat space BPS ` t Hooft-Polyakov monopole, obtained
from YM fields has a mass $4\pi v $, where  $v$ is the expectation 
value of $|A_0^a|$ at spatial
infinity. Even though the explicit solutions of the flat space and the curved-space BPS monopole are
quite different, they nevertheless allow similar interpretations in the non-singular gauge (\ref{duff}).
We have $ A_0 ( r \rightarrow \infty )  = 1/4M$ : the inverse of the Schwarzschild radius determines the
symmetry breaking scale  and  so the mass of the curved BPS monopole is
\be
M_{monopole} =  {\pi \over M} = 4\pi  A_0 ( r \rightarrow \infty  ) 
\ee
    
The second, dyon, solution reported in \cite{cd2} is 
\be
A_1 = \phi_1 = \phi_2 = 0, \hskip 1 cm A_0 =  \pm( c  - {1\over r})
\ee
Computing  (\ref{electmag}) for this solution  one observes 
that this solution has both 
electric and magnetic charges of unity.  Its abelian  nature was also explored in \cite{etesi} from a different perspective. 
The YM potential for this BPS dyon is twice that of the monopole (\ref{pot2}). 

Departing from CD, new solutions can be obtained, one of which is the following,~\footnote{Dyon solution
is $A_1 = \pm ( c - t/r^2)$ and the rest of the fields vanish.} 
\be
&&\tilde{A_0} = 0, \hskip 2 cm \tilde{A_1}=  - {2 M t\over r^3}, \cr 
&& \tilde{\phi_1} = 
(1 - {2M\over r})^{1\over 2}\cos ({ Mt\over r^2} +\omega_0) \hskip 1 cm
\tilde{\phi_2} = (1 - {2M\over r})^{1\over 2}\sin({Mt\over r^2} +\omega_0).
\label{weyl}
\ee 
This is again a topological charge 1 solution but since $\tilde{A_0^a} = 0$, it 
does not seem to allow a  monopole interpretation. 
But one can easily show that this solution is related to the CD solution (\ref{duff})
by a large gauge transformation $W(\vec{x},t ) = \mbox{exp}(-i f(r,t)\hat{x}\cdot \sigma/2)$, 
where  $f(r,t) =  t({M\over r^2}- \lambda )$. Therefore (\ref{weyl}) 
is simply in  ``bad'', Weyl, gauge
as far the monopole structure is concerned. The Yang-Mills potential and rest of the gauge 
invariant objects are the same as the CD solution of course. But the decomposition of the topological 
charge differs in these two gauges.

Following the discussion of \cite{dunne}, let us now see how the 
topological charge
can be  decomposed into radial part and and the Chern-Simons parts,
\begin{eqnarray}
Q &=& {1\over 8 \pi^2}\int_{M} \mbox{tr} F\wedge F \nonumber \\ 
  &=&  4\pi \int_{\Sigma} d^2 x \left\{ \epsilon_{\mu\nu} \epsilon_{ij} D_\mu \phi_i  D_\nu \phi_j
      -{1\over 2} \epsilon_{\mu \nu} F_{\mu \nu} (1- \phi_i\phi_i) \right \} \nonumber \\
 &=&\int_{0}^{8\pi M} dt \frac{d}{dt} CS(t) +\int_{2M}^\infty dr \frac{d}{dr} K(r).
\label{split}
\end{eqnarray}
In terms of the ansatz fields , the Chern-Simons term~\footnote{Our signs are different from those of \cite{dunne}}
$CS\equiv \int d^3x K_0$, is
\begin{eqnarray}
CS(t)= \frac{1}{2\pi} \int_{2M}^\infty dr\left[\phi_1^\prime \phi_2 -\phi_1
\phi_2^\prime -A_1(1-\phi_1^2-\phi_2^2) +\phi_1^\prime\right]
\label{cs}
\end{eqnarray}
while the radial part is
\begin{eqnarray}
K(r)=\frac{1}{2\pi} \int_{0}^{8\pi M} dt\left[\phi_1 \dot{\phi_2}
-\phi_2\dot{\phi_1} +A_0(1-\phi_1^2-\phi_2^2) -\dot{\phi_1}\right]
\label{kk}
\end{eqnarray}

For CD gauge (\ref{duff}), we have 
\be
CS(t) = 0 \hskip 1 cm K(r)= 1 - {8M^3\over r^3}, 
\label{cs2}
\ee
for which all the contribution to the topological charge comes from the radial part.
In the Weyl gauge (\ref{weyl}),
\be
CS(t) =  {t\over 8\pi M} \hskip 1 cm  K(r) = {4 M^2\over r^2} ( 1- {2 M\over r})
\label{cs3}
\ee
all the contribution to the topological charge comes form the Chern-Simons part.

In conclusion we have demonstrated that the apparently time periodic instanton solutions 
in Euclidean Schwarzschild background are actually time independent and do not describe 
tunneling.  To get this result, it was important to work in the proper gauge. All the 
solutions reported in the literature (including the ones we proposed here)  are monopoles 
in the sense of 't Hooft and Polyakov: they are non-singular three-dimensional solutions
with an associated abelian field strength.  This clarifies the 
physical interpretation of these solutions. 

We have also argued that even though the topological obstruction brought by gravity can be
circumvented, the dimensionful scale (Schwarzschild radius) 
brought by it eliminates four dimensional (arbitrary size) instantons. 

How about `constrained instantons' \cite{affleck} ?  Looking at the equations of motion (\ref{selfdual1})
we see that for $r >> 2M$, the equations reduce to those of flat space with a compact time. 
Therefore the usual caloron solution is \cite{hk} 
indeed a solution for $r >> 2 M $. 
\be 
A_0&=&- \partial_r \ln \rho, \hskip 2 cm  
A_1=\partial_0 \ln\rho\nonumber\\
\phi_1&=&r\,\partial_0 \ln\rho, \hskip 2cm  \phi_2=1-r\partial_r\ln\rho
\label{larger}
\ee 
where
\be
\rho(r,t)&=&1 +{\lambda^2 \over 4M r} {\sinh ( {r \over 4M}) 
\over \cosh ({r \over 4M})- \cos( {t\over 4M})}   
\label{finiterho}
\ee 
But clearly, for distances  $r >> \beta = 8\pi M$, time dependence drops out and the caloron
looks exactly like a 3-dimensional (time-independent) dipole \cite{gross}.  
Therefore because of the proximity of the two scales, $\beta$ and $2 M$, there is practically no
room for time dependent `constrained instanton' solutions.~\footnote{Here we exclusively deal with 
periodic solutions. On the other hand,  `quasi-periodic' solutions can be constructed for multi-instantons.
(A quasi-periodic solution broadly means that the ratio of the periods of two constituent 
instantons is irrational and one cannot define a `universal' period ). Quasi-periodic solutions in {\it{flat}}
space were found by Chakrabarti in \cite{chakrabarti}. It remains to be seen if Euclidean Schwarzschild 
space allows these type of solutions. } Numerical study of the self-duality 
equations would be necessary to decide this issue.

I am grateful to Stanley Deser for extended discussions, for helping me in every step of the 
work. I would like to thank Gerald Dunne for useful correspondence.
This work was supported by National Science Foundation  grant PHY99-73935

\myend
\begin{thebibliography}{99}

\bibitem{cd1}
J.~M.~Charap and M.~J.~Duff,
Phys.\ Lett.\ B {\bf 69}, (1977) 445.

\bibitem{cd2}
J.~M.~Charap and M.~J.~Duff,
Phys.\ Lett.\ B {\bf 71}, (1977) 219 .


\bibitem{polyakov}
A.~A.~Belavin, A.~M.~Polyakov, A.~S.~Schwarts and Y.~S.~Tyupkin,
Phys.\ Lett.\ B {\bf 59} (1975) 85.


\bibitem{bk}
R.~Bartnik and J.~McKinnon,
Phys.\ Rev.\ Lett.\  {\bf 61}, (1988) 141


\bibitem{deser1}
S.~Deser,
Phys.\ Lett.\ B {\bf 64}, 463 (1976).


\bibitem{ferstl}
A.~Ferstl, B.~Tekin and V.~Weir,
Phys.\ Rev.\ D {\bf 62}, (2000) 064003
[arXiv:hep-th/0002019].



\bibitem{chak}
H.~Boutaleb-Joutei, A.~Chakrabarti and A.~Comtet,
Phys.\ Rev.\ D {\bf 20}, 1898 (1979).

\bibitem{hk}
B.~J.~Harrington and H.~K.~Shepard,
Phys.\ Rev.\ D {\bf 17}, 2122 (1978).


\bibitem{dunne}
G.~V.~Dunne and B.~Tekin,
Phys.\ Rev.\ D {\bf 63}, 085004 (2001)
[arXiv:hep-th/0011169].



\bibitem{hawking}
S.~W.~Hawking,
Phys.\ Lett.\ A {\bf 60}, 81 (1977).


\bibitem{pisarski}
R.~D.~Pisarski,
Phys.\ Rev.\ D {\bf 35}, 664 (1987).


\bibitem{gerald}
F.~T.~Brandt, A.~Das, G.~V.~Dunne, J.~Frenkel and J.~C.~Taylor,
arXiv:hep-th/0111146.


\bibitem{affleck}
I.~Affleck,
Nucl.\ Phys.\ B {\bf 191}, 429 (1981).

\bibitem{witten}
E.~Witten,
Phys.\ Rev.\ Lett.\  {\bf 38}, 121 (1977).


\bibitem{hosotani}
J.~Bjoraker and Y.~Hosotani,
Phys.\ Rev.\ D {\bf 62}, 043513 (2000)
[arXiv:hep-th/0002098].


\bibitem{thooft}
G.~'t Hooft,
Nucl.\ Phys.\ B {\bf 79}, 276 (1974).


\bibitem{ad}
L.~F.~Abbott and S.~Deser,
Phys.\ Lett.\ B {\bf 116}, 259 (1982).


\bibitem{etesi}
G.~Etesi and T.~Hausel,
J.\ Geom.\ Phys.\  {\bf 37}, 126 (2001)
[arXiv:hep-th/0003239].

\bibitem{comtet}
A.~Comtet,
Phys.\ Rev.\ D {\bf 18},(1978) 3890

\bibitem{tekin}
B.~Tekin,
JHEP {\bf 0008},(2000) 049
[arXiv:hep-th/0006135].

\bibitem{barti}
H.~Boutaleb-Joutei, A.~Chakrabarti and A.~Comtet,
Phys.\ Rev.\ D {\bf 20}, 1884 (1979).

\bibitem{gross}
D.~J.~Gross, R.~D.~Pisarski and L.~G.~Yaffe,
Rev.\ Mod.\ Phys.\  {\bf 53}, 43 (1981).

\bibitem{chakrabarti}
A.~Chakrabarti,
Phys.\ Rev.\ D {\bf 38}, 3219 (1988).


\end{thebibliography}
